\documentclass[conference]{IEEEtran}
\usepackage{hyperref}
\usepackage[noadjust]{cite}
\usepackage[cmex10]{amsmath}
\usepackage{esint}
\usepackage{commath}
\usepackage[caption=false,font=footnotesize]{subfig}
\usepackage{url}
\usepackage{graphicx}
\usepackage{amssymb}
\everymath{\displaystyle}
\usepackage{comment}

\usepackage[shortlabels]{enumitem}
\usepackage{nameref}
\usepackage{slashbox}
\usepackage{textcomp}
\usepackage{gensymb}
\usepackage{tikz}
\usepackage{tikz-3dplot}
\usetikzlibrary{patterns}
\usepackage{tkz-euclide}

\DeclareMathOperator{\floor}{floor}

\mathchardef\UrlBreakPenalty=10000
\mathchardef\UrlBigBreakPenalty=10000

\onecolumn

\begin{document}
\title
{
  Efficient Kernel Convolution\\
  for Smooth Surfaces without Edge Effects
}
\author
{
\IEEEauthorblockN{Alexander Gribov}
\IEEEauthorblockA
{
  Environmental Systems Research Institute\\
  380 New York Street\\
  Redlands, CA 92373\\
  E-mail: agribov@esri.com}
}

\maketitle

\begin{abstract}
\boldmath
One of the most efficient ways to produce unconditional simulations is with the kernel convolution using fast Fourier transform (FFT) \cite{KernelConvolutionFFT}. However, when data is located on a surface, this approach is not efficient because data needs to be processed in a three-dimensional enclosing box. This paper describes a novel approach based on integer transformation to reduce the volume of the enclosing box.
\end{abstract}


\section{Introduction}

The purpose of this paper is to develop an approach to efficiently apply kernel convolution using FFT \cite{KernelConvolutionFFT} to smooth surfaces (sphere, ellipsoid, etc.). Let's define a cubical lattice covering a three-dimensional space. Each point on the surface can be represented as a weighted combination of nodes or randomly assigned to one node from a corresponding lattice cell. Note that errors are introduced in this step. The convolution can be performed in three-dimensional space; however, it will not be computationally efficient. Even dividing data by tiles and processing them separately is not efficient because the data in each tile only occupies a slice. Another approach would be to rotate each tile to reduce the volume of the enclosing box. However, this solution will not maintain nodes and, therefore, will lead to an edge effect. The edge effect might be resolved by using overlapping tiles and, after applying convolution, merging the solutions.

The next section will describe an approach based on integer transformation that leads to a computationally efficient method and eliminates the edge effect.

\section{Approach Based on Integer Transformation}

Because the complexity of kernel convolution using FFT depends on the volume of the enclosing box, the idea is to find a linear transformation for the set of nodes that reduces its volume. To maintain the lattice structure, the transformation matrix should be an integer. Let $A$ be a three-by-three full rank matrix that has integer components. Further in the text, the word ``integer'' will be often omitted. Note that the matrix is not necessarily orthogonal. Using transformation matrix $A$, which minimizes the volume of the enclosing box, will significantly reduce the complexity of kernel convolution using FFT. This solution does not have any edge effect because all original nodes have a corresponding transformed node. However, only in the case where the absolute value of the determinant of $ A $ is equal to $ 1 $, every transformed node has a corresponding original node. The transformation is also applied to the kernel. The transformed kernel can be found by back transformation using the inverse matrix.

\begin{figure}
  \centering
  \begin{tikzpicture} [scale = 0.9]
    \tkzInit[xmin = 0, ymin = 0, xmax = 14, ymax = 22]
    \tkzGrid
    \tkzAxeXY

    \draw [ultra thick, green] ( 1,  1) -- ( 1,  2) -- ( 5,  9) -- (13, 21) -- ( 9, 13) -- cycle;

    \fill [green] ( 1,  1) circle [radius = 5pt];
    \fill [green] ( 1,  2) circle [radius = 5pt];
    \fill [green] ( 5,  9) circle [radius = 5pt];
    \fill [green] (13, 21) circle [radius = 5pt];
    \fill [green] ( 9, 13) circle [radius = 5pt];

    \fill ( 1,  1) circle [radius = 3pt];
    \fill ( 3,  4) circle [radius = 3pt];
    \fill ( 5,  7) circle [radius = 3pt];
    \fill ( 7, 10) circle [radius = 3pt];
    \fill ( 9, 13) circle [radius = 3pt];
    \fill ( 2,  3) circle [radius = 3pt];
    \fill ( 4,  6) circle [radius = 3pt];
    \fill ( 6,  9) circle [radius = 3pt];
    \fill ( 8, 12) circle [radius = 3pt];
    \fill (10, 15) circle [radius = 3pt];
    \fill ( 3,  5) circle [radius = 3pt];
    \fill ( 5,  8) circle [radius = 3pt];
    \fill ( 7, 11) circle [radius = 3pt];
    \fill ( 9, 14) circle [radius = 3pt];
    \fill (11, 17) circle [radius = 3pt];
    \fill ( 4,  7) circle [radius = 3pt];
    \fill ( 6, 10) circle [radius = 3pt];
    \fill ( 8, 13) circle [radius = 3pt];
    \fill (10, 16) circle [radius = 3pt];
    \fill (12, 19) circle [radius = 3pt];
    \fill ( 5,  9) circle [radius = 3pt];
    \fill ( 7, 12) circle [radius = 3pt];
    \fill ( 9, 15) circle [radius = 3pt];
    \fill (11, 18) circle [radius = 3pt];
    \fill (13, 21) circle [radius = 3pt];
    \fill ( 1,  2) circle [radius = 3pt];
  \end{tikzpicture}
  \caption
  {
    Convex hull (green polygon) around source data (black points). Large green points are vertices of the convex hull.
  }
  \label{fig:TwoDimensionalExampleConvexHull}
\end{figure}
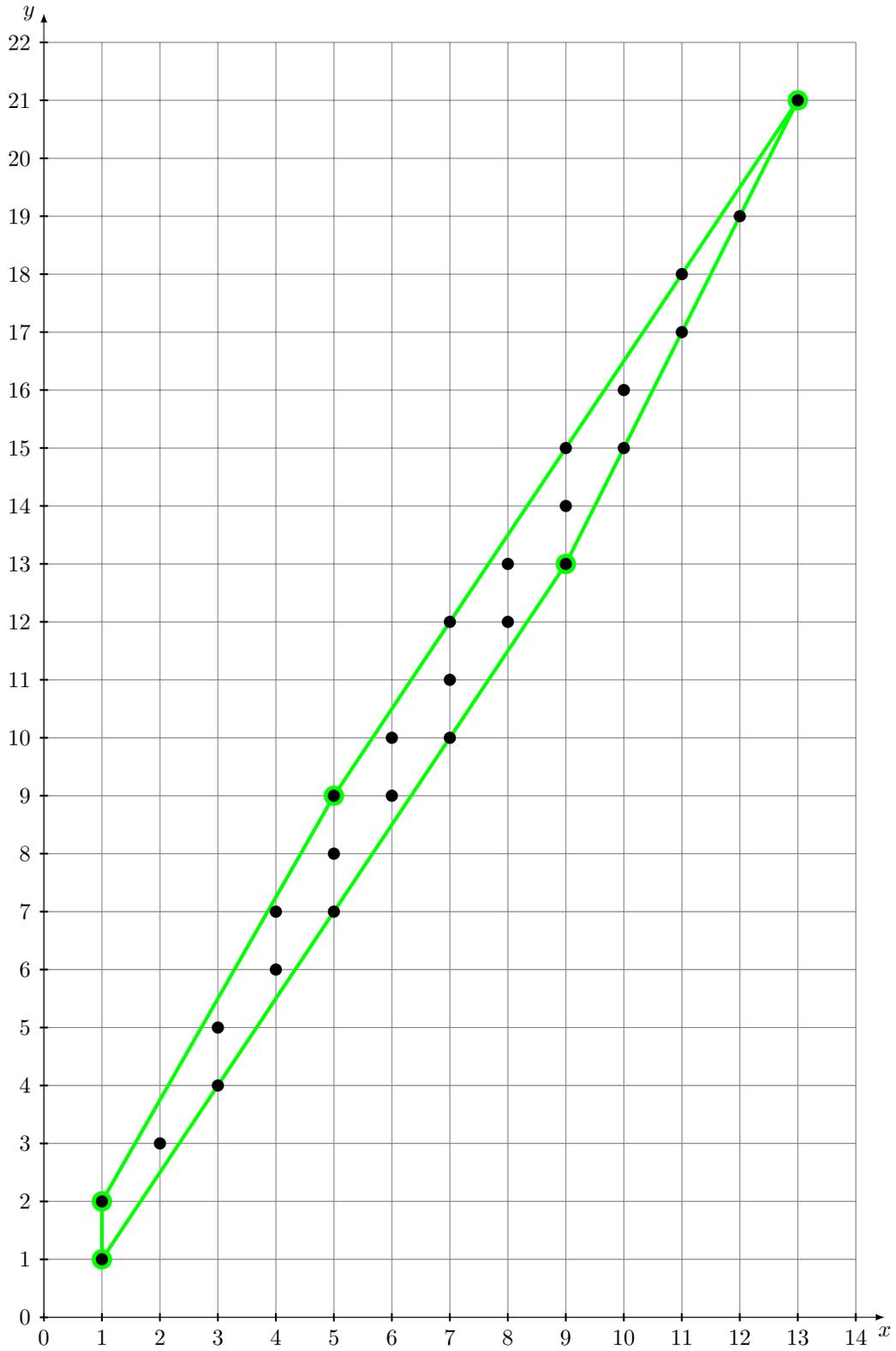

Each row of the matrix $A$ defines a vector for the scalar projection of the source data. To minimize the enclosing box, each vector should project source data to the smallest interval. The only requirement is that the vectors are linearly independent. The algorithm to find a set of linearly independent vectors, which minimizes intervals, will be explained by a two-dimensional example. The three-dimensional case is similar. Source data is shown in Figure~\ref{fig:TwoDimensionalExampleConvexHull}.

\begin{enumerate}
  \item
  {
    Construct a convex hull around the source data (Figure~\ref{fig:TwoDimensionalExampleConvexHull}).
  }
  \item
  {
    Find antipodal points (Figure~\ref{fig:TwoDimensionalExampleDirections}). In the two-dimensional case antipodal points are found by rotating calipers.

    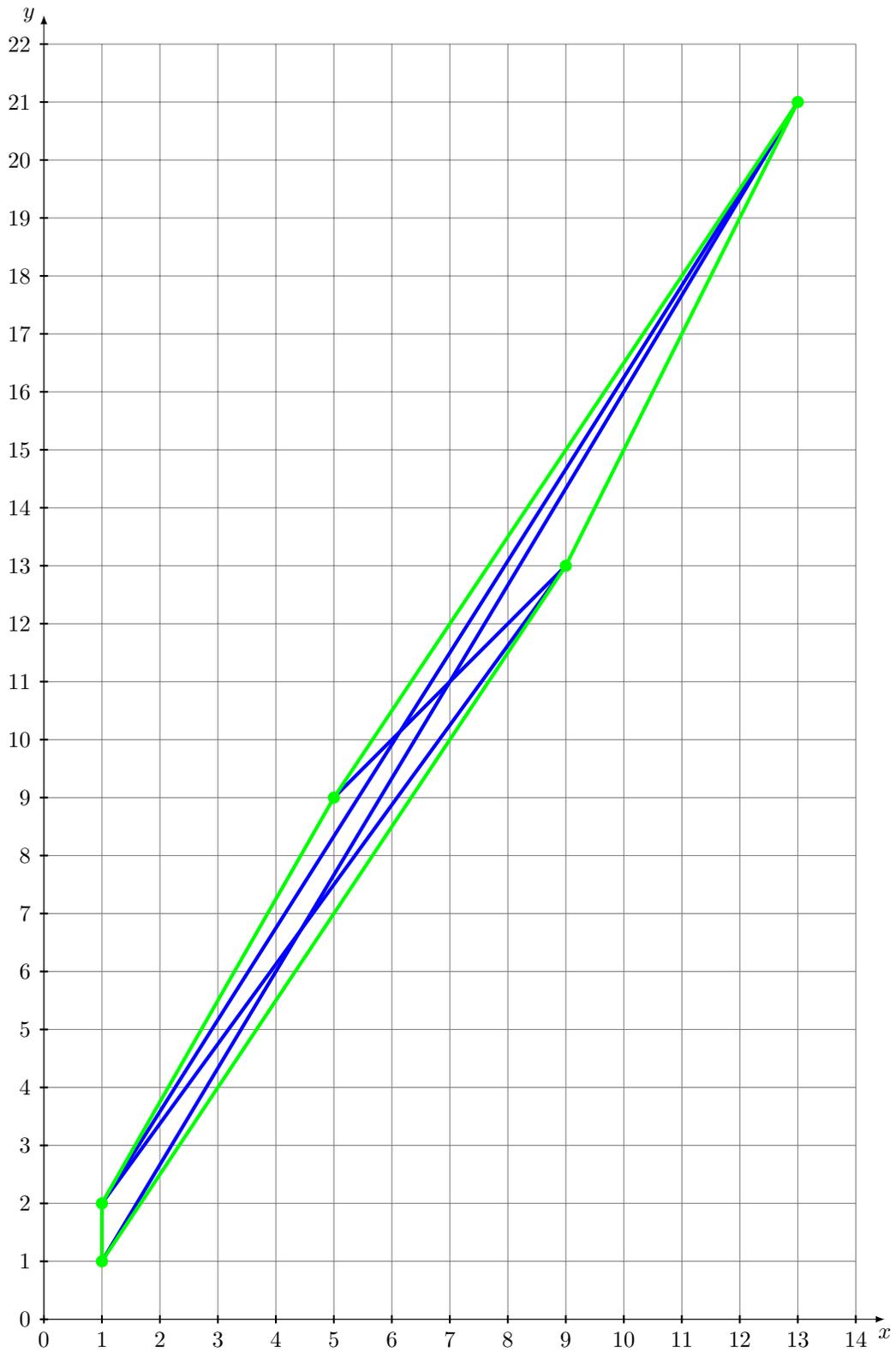
\begin{figure}
      \centering
      \begin{tikzpicture} [scale = 0.9]
        \tkzInit[xmin = 0, ymin = 0, xmax = 14, ymax = 22]
        \tkzGrid
        \tkzAxeXY
    
        \draw [ultra thick, blue] ( 1,  1) -- (13, 21);
        \draw [ultra thick, blue] ( 1,  2) -- (13, 21);
        \draw [ultra thick, blue] ( 1,  2) -- ( 9, 13);
        \draw [ultra thick, blue] ( 5,  9) -- ( 9, 13);
    
        \draw [ultra thick, green] ( 1,  1) -- ( 1,  2) -- ( 5,  9) -- (13, 21) -- ( 9, 13) -- cycle;
    
        \fill [green] ( 1,  1) circle [radius = 3pt];
        \fill [green] ( 1,  2) circle [radius = 3pt];
        \fill [green] ( 5,  9) circle [radius = 3pt];
        \fill [green] (13, 21) circle [radius = 3pt];
        \fill [green] ( 9, 13) circle [radius = 3pt];
      \end{tikzpicture}
      \caption
      {
        Antipodal points (endpoints of the blue segments) of the convex hull (green polygon).
      }
      \label{fig:TwoDimensionalExampleDirections}
    \end{figure}
  }
  \item
  {
    Each pair of antipodal points is related to the spread of its rotating calipers (Figure~\ref{fig:TwoDimensionalExampleVectors}). All antipodal points and vectors perpendicular to rotating calipers are shown in Figure~\ref{fig:TwoDimensionalExampleAllVectors}. Each vector is divided by the greatest common divisor of all the coordinates.

    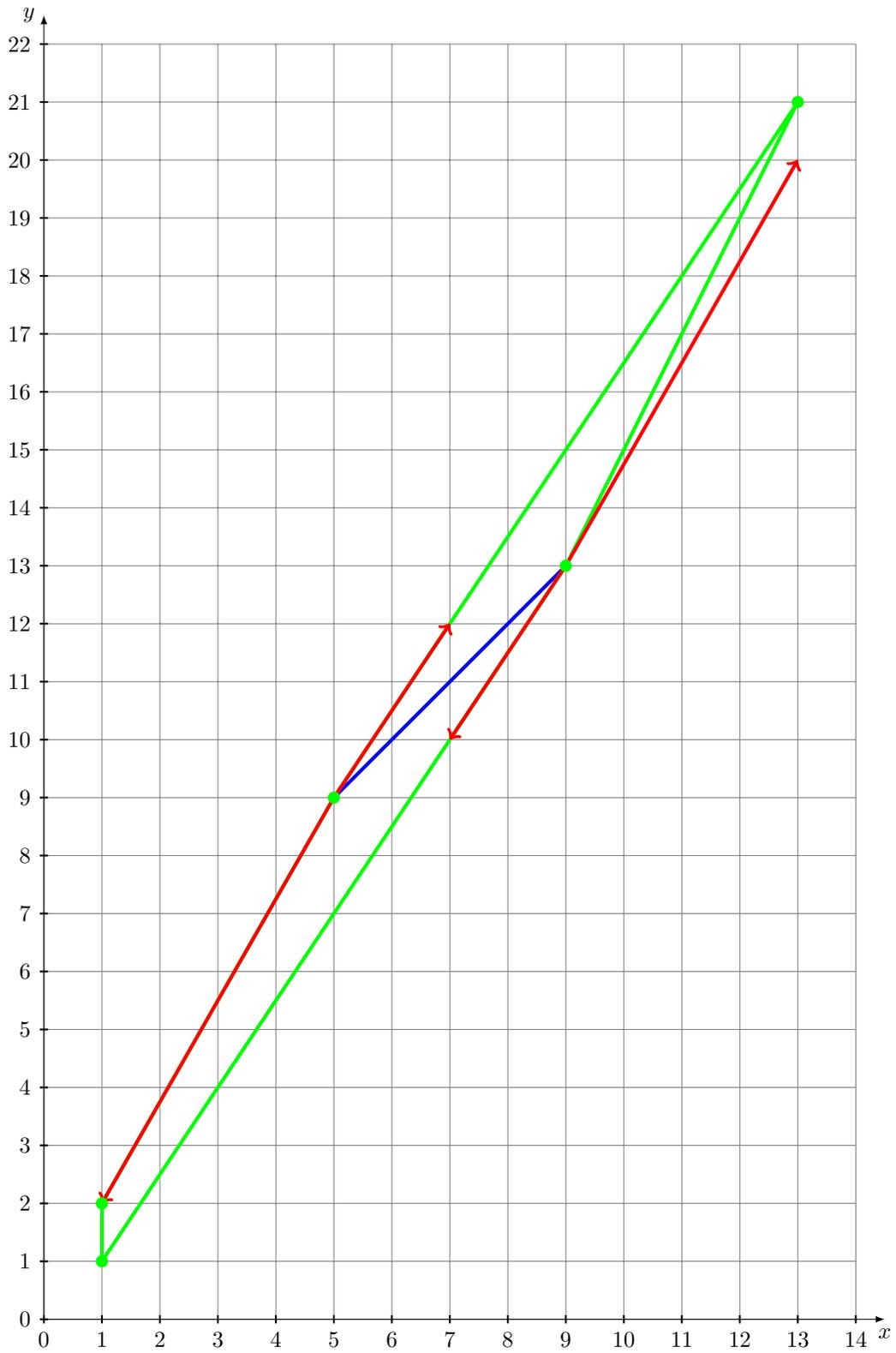
\begin{figure}
      \centering
      \begin{tikzpicture} [scale = 0.9]
        \tkzInit[xmin = 0, ymin = 0, xmax = 14, ymax = 22]
        \tkzGrid
        \tkzAxeXY

        \draw [ultra thick, blue] ( 5,  9) -- ( 9, 13);

        \draw [ultra thick, green] ( 1,  1) -- ( 1,  2) -- ( 5,  9) -- (13, 21) -- ( 9, 13) -- cycle;

        \draw [->, ultra thick, red] ( 5,  9) -- ( 1,  2);
        \draw [->, ultra thick, red] ( 5,  9) -- ( 7, 12);
        \draw [->, ultra thick, red] ( 9, 13) -- (13, 20);
        \draw [->, ultra thick, red] ( 9, 13) -- ( 7, 10);

        \fill [green] ( 1,  1) circle [radius = 3pt];
        \fill [green] ( 1,  2) circle [radius = 3pt];
        \fill [green] ( 5,  9) circle [radius = 3pt];
        \fill [green] (13, 21) circle [radius = 3pt];
        \fill [green] ( 9, 13) circle [radius = 3pt];
      \end{tikzpicture}
      \caption
      {
        Rotating calipers (red vectors) for the pair of antipodal points (endpoints of the blue segment).
      }
      \label{fig:TwoDimensionalExampleVectors}
    \end{figure}

    \begin{figure}
      \centering
      \begin{tabular}{c c}
        \begin{tikzpicture} [scale = 0.45]
          \tkzInit[xmin = -4, ymin = -1, xmax = 13, ymax = 21]
          \tkzGrid
          \tkzAxeXY
      
          \draw [->, ultra thick, blue] ( 0,  0) -- (12, 20);
          \draw [->, ultra thick, red] ( 0,  0) -- (-3,  2);
          \draw [->, ultra thick, red] ( 0,  0) -- ( 1,  0);
        \end{tikzpicture}
        &
        \begin{tikzpicture} [scale = 0.45]
          \tkzInit[xmin = -1, ymin = -2, xmax = 13, ymax = 20]
          \tkzGrid
          \tkzAxeXY
      
          \draw [->, ultra thick, blue] ( 0,  0) -- (12, 19);
          \draw [->, ultra thick, red] ( 0,  0) -- ( 1,  0);
          \draw [->, ultra thick, red] ( 0,  0) -- ( 2, -1);
        \end{tikzpicture}
        \\
        a)
        &
        b)
        \\
        \begin{tikzpicture} [scale = 0.45]
          \tkzInit[xmin = -1, ymin = -5, xmax = 9, ymax = 12]
          \tkzGrid
          \tkzAxeXY
      
          \draw [->, ultra thick, blue] ( 0,  0) -- ( 8, 11);
          \draw [->, ultra thick, red] ( 0,  0) -- ( 2, -1);
          \draw [->, ultra thick, red] ( 0,  0) -- ( 7, -4);
        \end{tikzpicture}
        &
        \begin{tikzpicture} [scale = 0.45]
          \tkzInit[xmin = -1, ymin = -5, xmax = 8, ymax = 5]
          \tkzGrid
          \tkzAxeXY
      
          \draw [->, ultra thick, blue] ( 0,  0) -- ( 4,  4);
          \draw [->, ultra thick, red] ( 0,  0) -- ( 7, -4);
          \draw [->, ultra thick, red] ( 0,  0) -- ( 3, -2);
        \end{tikzpicture}
        \\
        c)
        &
        d)
      \end{tabular}
      \caption
      {
        Set of all antipodal points for the convex hull (see Figure~\ref{fig:TwoDimensionalExampleDirections}) with rotating calipers (perpendicular to red vectors).
      }
      \label{fig:TwoDimensionalExampleAllVectors}
    \end{figure}
  }
  \item
  {
    It is sufficient to check all vectors inside parallelograms and on the edges of parallelograms with the exception of the origin of the coordinate system and opposite vertices for each parallelogram (Figure~\ref{fig:TwoDimensionalExampleParallelogramsContainingAllRegions}).

    Let's prove this statement for any number of dimensions by contradiction. Suppose that the optimal set of the projection vectors has at least one vector $ \vec{v} $ outside its parallelepiped or at the vertex opposite the origin of the coordinate system. Let $ \vec{v}_i $ be vectors defining the parallelepiped. Then
    \begin{equation*}
      \vec{v}
      =
      \sum
      {
        \vec{v}_i
        w_i
      }
      ,
    \end{equation*}
    where $ w_i $ are weights, $ 0 \leq w_i $.
    Because $ \vec{v} $ is outside the parallelepiped, one of its weights should be greater than $ 1 $. Suppose it has index $ i^{\star} $. Replace that weight by its fractional part and obtain a new vector $ \vec{u} $. $ \vec{v}_{i^{\star}} $ and $ \vec{u} $ will have a smaller projection than $ \vec{v} $. If they are both collinear to other projection vectors, then $ \vec{v} $ is also collinear. Therefore, one of them can be used instead of $ \vec{v} $. This is not possible, because the original set of projection vectors is optimal. The case where all weights are equal to $ 1 $ can also be skipped because all $ \vec{v}_i$ will not have a larger projection and will form a complete basis. Therefore, one of them should not be collinear to other projection vectors. This proves that it is sufficient to analyze vectors inside the parallelepiped and vectors used to form it.

    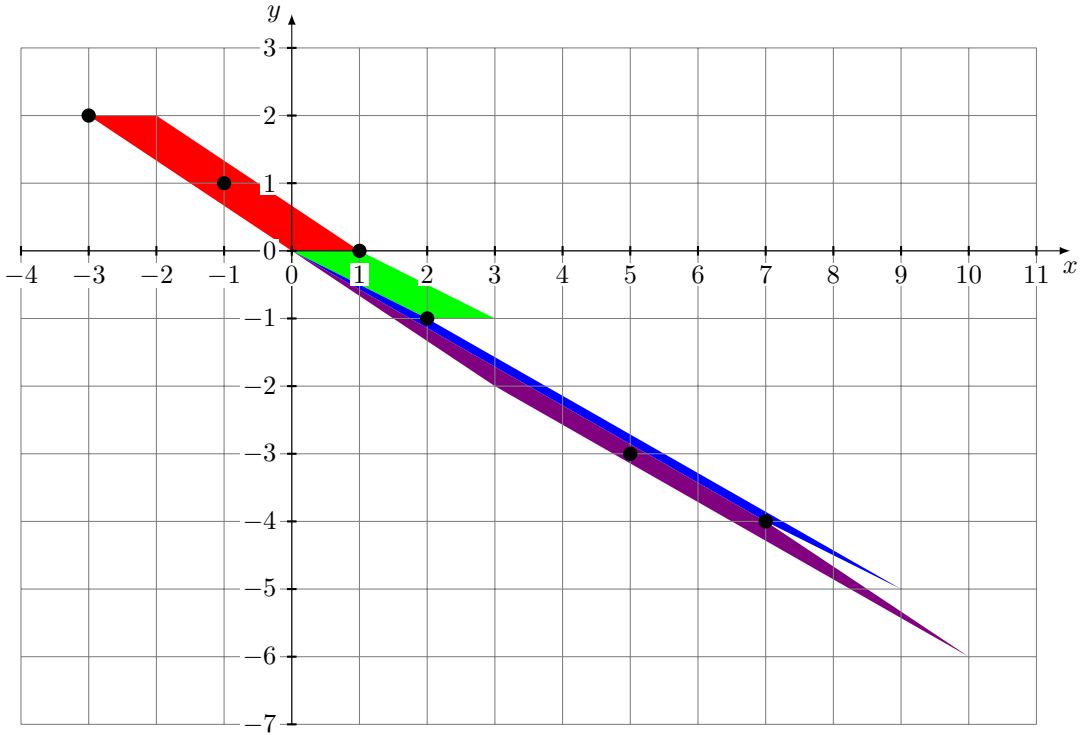
\begin{figure}
      \centering
      \begin{tikzpicture} [scale = 0.9]
        \tkzInit[xmin = -4, ymin = -7, xmax = 11, ymax = 3]
    
        \fill [red] ( 0,  0) -- (-3,  2) -- (-2,  2) -- ( 1,  0) -- cycle;
    
        \fill [green] ( 0,  0) -- ( 1,  0) -- ( 3, -1) -- ( 2, -1) -- cycle;
    
        \fill [blue] ( 0,  0) -- ( 2, -1) -- ( 9, -5) -- ( 7, -4) -- cycle;
    
        \fill [violet] ( 0,  0) -- ( 7, -4) -- (10, -6) -- ( 3, -2) -- cycle;
    
        \tkzGrid
        \tkzAxeXY
    
        \fill (-3,  2) circle [radius = 3pt];
        \fill ( 1,  0) circle [radius = 3pt];
        \fill ( 2, -1) circle [radius = 3pt];
        \fill ( 7, -4) circle [radius = 3pt];
    
        \fill (-1,  1) circle [radius = 3pt];
        \fill ( 5, -3) circle [radius = 3pt];
      \end{tikzpicture}
      \caption
      {
        Parallelograms (red, blue, green, violet) obtained by perpendiculars of rotating calipers. It is sufficient to consider only the black points corresponding to a set of vectors.
      }
      \label{fig:TwoDimensionalExampleParallelogramsContainingAllRegions}
    \end{figure}

    To find all points inside the parallelogram, it is divided into two triangles by the diagonal not touching the origin of the coordinate system. Then, for each triangle, if it has a point on any of its edges, it is divided into two triangles (see \nameref{sec:FindingAnIntegerPointInsideAnIntegerTriangle}). If a triangle does not have any points on its edges but has point inside, it is divided into three triangles (see \nameref{sec:FindingAnIntegerPointInsideAnIntegerTriangle}).

    If any point on the edges of or inside a triangle cannot improve the already found solution, then such a triangle can be omitted from analysis. This is tested by checking all vertices of the triangle against the pair of antipodal points. By construction, this pair is the same for all points of the triangle.
  }
\end{enumerate}

Next, two linearly independent vectors will have the smallest intervals: $\left( 2, -1 \right)$ and $\left( -3, 2 \right)$. This corresponds to the transformation matrix
\begin{equation}
  \begin{bmatrix}
     2 & -1\\
    -3 &  2
  \end{bmatrix}
  .
  \label{eq:TwoDimensionalExampleIntegerTransformationMatrix}
\end{equation}

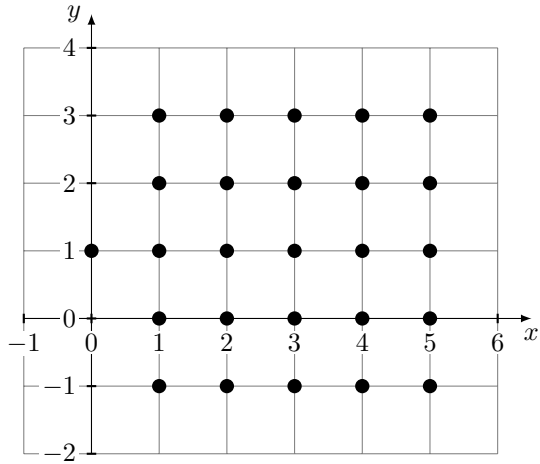
\begin{figure}
  \centering
  \begin{tikzpicture} [scale = 0.9]
    \tkzInit[xmin = -1, ymin = -2, xmax = 6, ymax = 4]
    \tkzGrid
    \tkzAxeXY

    \fill (1, -1) circle [radius = 3pt];
    \fill (2, -1) circle [radius = 3pt];
    \fill (3, -1) circle [radius = 3pt];
    \fill (4, -1) circle [radius = 3pt];
    \fill (5, -1) circle [radius = 3pt];
    \fill (1,  0) circle [radius = 3pt];
    \fill (2,  0) circle [radius = 3pt];
    \fill (3,  0) circle [radius = 3pt];
    \fill (4,  0) circle [radius = 3pt];
    \fill (5,  0) circle [radius = 3pt];
    \fill (1,  1) circle [radius = 3pt];
    \fill (2,  1) circle [radius = 3pt];
    \fill (3,  1) circle [radius = 3pt];
    \fill (4,  1) circle [radius = 3pt];
    \fill (5,  1) circle [radius = 3pt];
    \fill (1,  2) circle [radius = 3pt];
    \fill (2,  2) circle [radius = 3pt];
    \fill (3,  2) circle [radius = 3pt];
    \fill (4,  2) circle [radius = 3pt];
    \fill (5,  2) circle [radius = 3pt];
    \fill (1,  3) circle [radius = 3pt];
    \fill (2,  3) circle [radius = 3pt];
    \fill (3,  3) circle [radius = 3pt];
    \fill (4,  3) circle [radius = 3pt];
    \fill (5,  3) circle [radius = 3pt];
    \fill (0,  1) circle [radius = 3pt];
  \end{tikzpicture}
  \caption
  {
    Transformed data from Figure~\ref{fig:TwoDimensionalExampleConvexHull} by the transformation matrix \eqref{eq:TwoDimensionalExampleIntegerTransformationMatrix}.
  }
  \label{fig:TwoDimensionalExampleTransformed}
\end{figure}

Figure~\ref{fig:TwoDimensionalExampleTransformed} shows the result of applying the transformation matrix \eqref{eq:TwoDimensionalExampleIntegerTransformationMatrix} to the source data shown in Figure~\ref{fig:TwoDimensionalExampleConvexHull}. The size of the bounding box changed from $273 = 13 \times 21$ to $30 = 6 \times 5$.

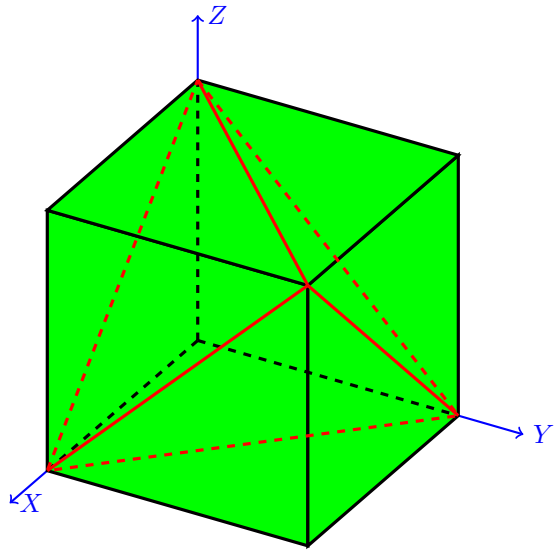
\begin{figure}
  \centering
  \tdplotsetmaincoords{60}{120}
  \begin{tikzpicture} [tdplot_main_coords, cube/.style={very thick, fill = green}, cube hidden/.style={very thick, dashed}, line/.style={very thick, red}, line hidden/.style={very thick, red, dashed}, grid/.style={very thin, gray}, axis/.style = {->, blue, thick}]
    \draw [axis] (0, 0, 0) -- (5, 0, 0) node [anchor = west] {$X$};
    \draw [axis] (0, 0, 0) -- (0, 5, 0) node [anchor = west] {$Y$};
    \draw [axis] (0, 0, 0) -- (0, 0, 5) node [anchor = west] {$Z$};

    \draw [cube] (4, 0, 0) -- (4, 4, 0) -- (4, 4, 4) -- (4, 0, 4) -- cycle;
    \draw [cube] (0, 4, 0) -- (4, 4, 0) -- (4, 4, 4) -- (0, 4, 4) -- cycle;
    \draw [cube] (0, 0, 4) -- (0, 4, 4) -- (4, 4, 4) -- (4, 0, 4) -- cycle;
    
    \draw [cube hidden] (0, 0, 0) -- (4, 0, 0);
    \draw [cube hidden] (0, 0, 0) -- (0, 4, 0);
    \draw [cube hidden] (0, 0, 0) -- (0, 0, 4);

    \draw [line hidden] (0, 0, 4) -- (0, 4, 0) -- (4, 0, 0) -- cycle;
    \draw [line] (0, 0, 4) -- (4, 4, 4);
    \draw [line] (0, 4, 0) -- (4, 4, 4);
    \draw [line] (4, 0, 0) -- (4, 4, 4);
  \end{tikzpicture}
  \caption
  {
    Dividing a parallelepiped into five tetrahedrons.
  }
  \label{fig:ParallelepipedToTetrahedrons}
\end{figure}

The algorithm described above is valid in higher dimensions. For a three-dimensional case, a parallelepiped can be divided into five tetrahedrons by the scheme shown in Figure~\ref{fig:ParallelepipedToTetrahedrons}.

The algorithm to further divide tetrahedrons is described in \nameref{sec:FindingAnIntegerPointInsideAnIntegerTetrahedron}.

\section{Three-Dimensional Example}

\begin{figure}
  \centering
  \begin{tabular}{c c c}
    \includegraphics[width = 8.75 cm, keepaspectratio]{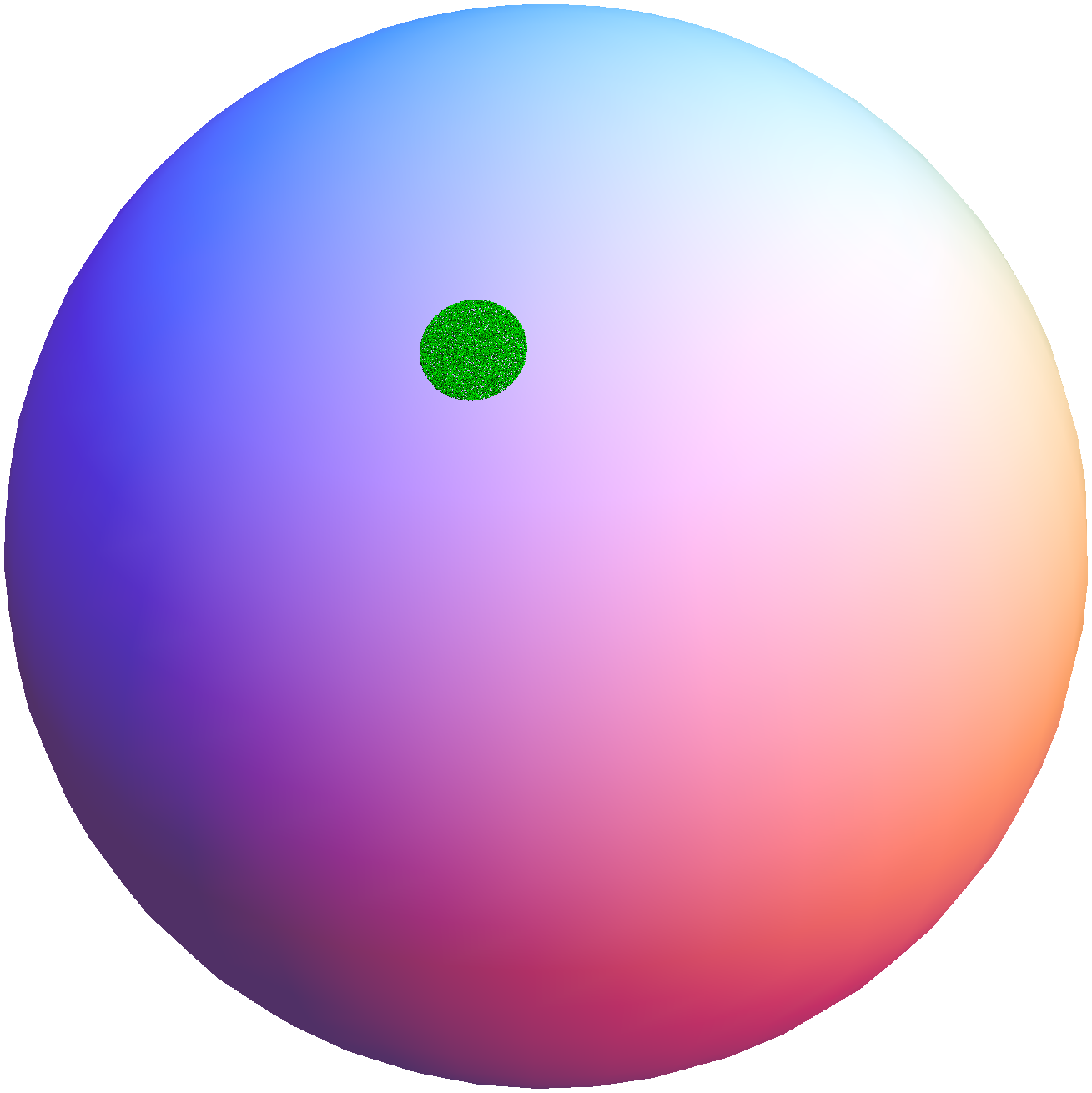}
    &
    \includegraphics[width = 8.75 cm, keepaspectratio]{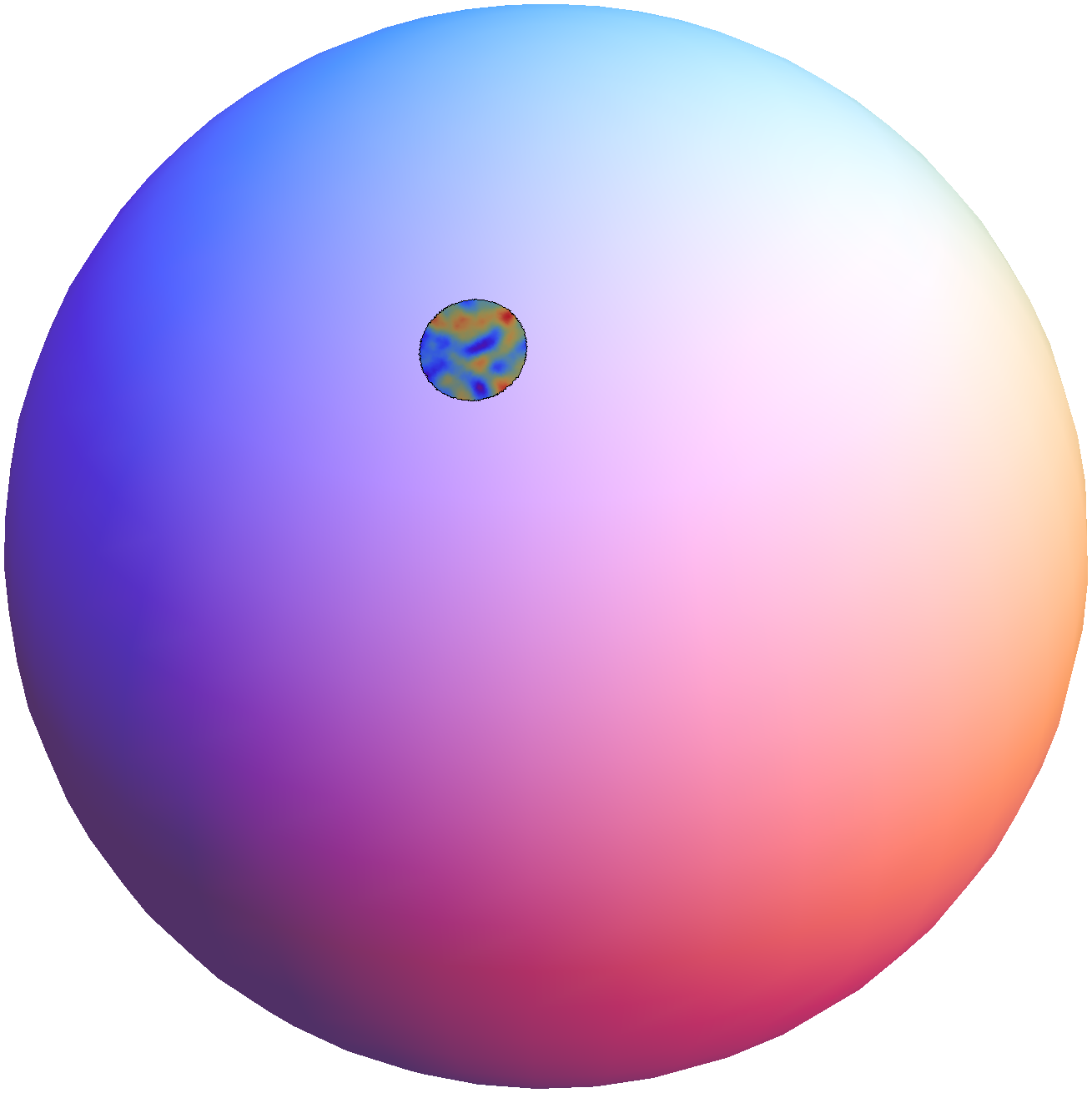}
    \\
    a)
    &
    b)
  \end{tabular}
  \caption
  {
    a) Location of points on the sphere.
    b) Unconditional simulations obtained by applying kernel $1 - \dfrac{h}{0.025}$ with bandwidth $0.025$.
  }
  \label{fig:SourceDataAndUnconditionalSimulations}
\end{figure}

\begin{figure}
  \centering
  \includegraphics[width = \columnwidth, keepaspectratio]{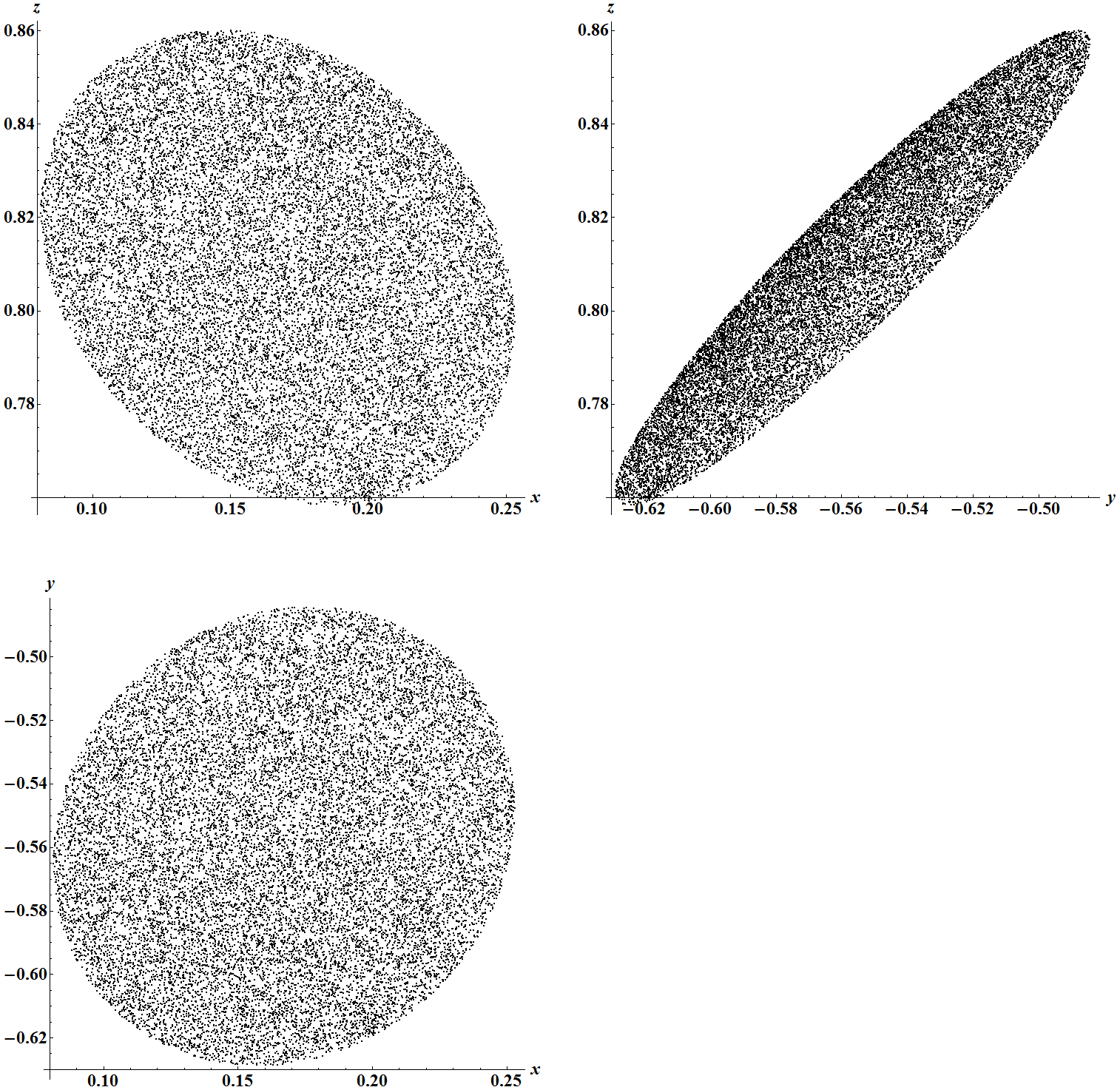}
  \caption
  {
    Points from Figure~\ref{fig:SourceDataAndUnconditionalSimulations}a are shown in three projections.
  }
  \label{fig:SourceDrawingProjections}
\end{figure}

The efficiency of the transformation is demonstrated on $20,000$ uniformly distributed random points in the spherical circle\footnote{A spherical circle is defined as a set of all points on the sphere surface having a central angle from the spherical circle center that is less than the spherical circle radius.} with the center $\left( 0.167775, -0.558644, 0.812261 \right)$ and the radius $5 \degree$. They are shown in Figure~\ref{fig:SourceDataAndUnconditionalSimulations}a, and their projections are shown in Figure~\ref{fig:SourceDrawingProjections}.

\begin{figure}
  \centering
  \includegraphics[width = \columnwidth, keepaspectratio]{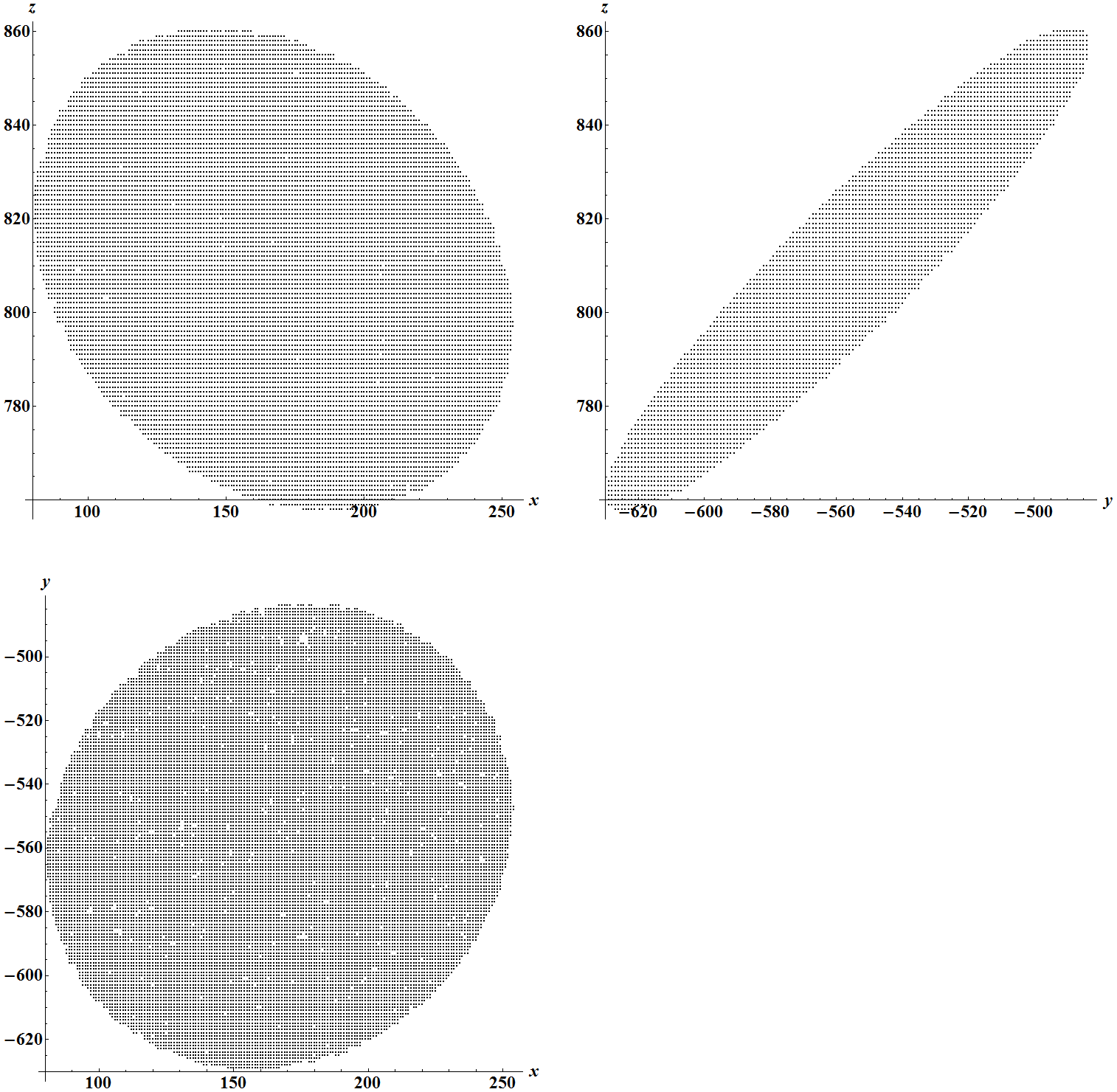}
  \caption
  {
    Weighted points on a three-dimensional lattice are shown in three projections.
  }
  \label{fig:SourceDrawingLatticeProjections}
\end{figure}

A three-dimensional lattice with a step size equal to $10^{-3}$ was defined. Each point was represented by weighted combinations of corresponding cell nodes (see Figure~\ref{fig:SourceDrawingLatticeProjections}).

\begin{figure}
  \centering
  \includegraphics[width = \columnwidth, keepaspectratio]{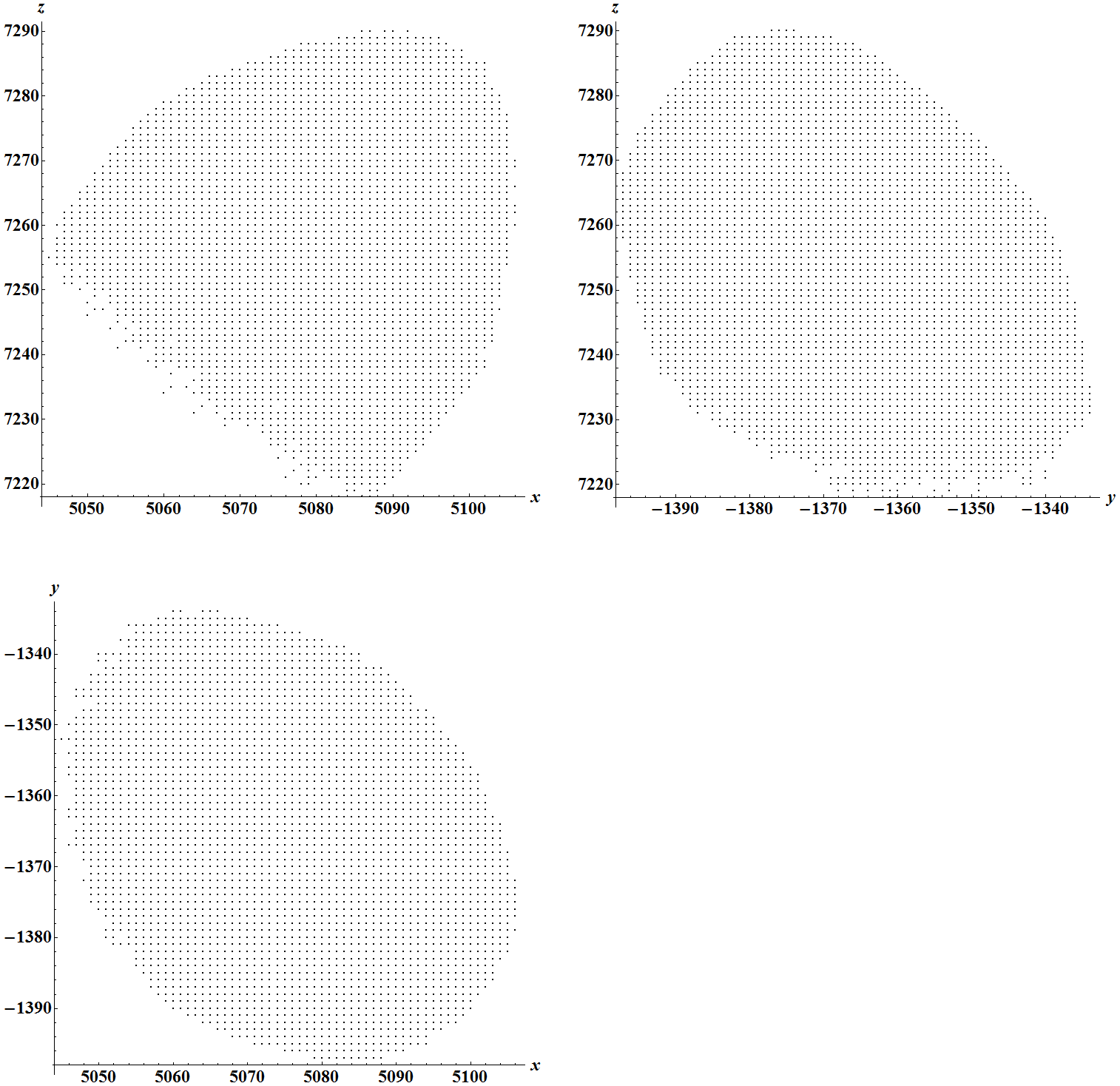}
  \caption
  {
    Transformed points from Figure~\ref{fig:SourceDrawingLatticeProjections} are shown in three projections.
  }
  \label{fig:SourceDrawingLatticeTransformedProjections}
\end{figure}

After transforming the nodes using the following transformation matrix,
\begin{equation*}
  \begin{bmatrix}
    1 & -3 &  4\\
    0 &  1 & -1\\
    1 & -4 &  6
  \end{bmatrix}
  ,
\end{equation*}
the projected points are shown in Figure~\ref{fig:SourceDrawingLatticeTransformedProjections}.

The enclosing box for the lattice points has sizes $174$, $146$, and $103$. After the transformation, the resultant enclosing box has sizes $62$, $64$, and $73$. This is about $9$ times smaller. Performing kernel convolution using FFT will be more than $9$ times faster due to $O{\left( n \cdot \log{\left( n \right)} \right)}$ complexity. Expected time improvement for tiles with a smaller spherical circular radius is even greater.

Figure~\ref{fig:SourceDataAndUnconditionalSimulations}b shows the result of kernel convolution using FFT.

Nonstationary unconditional simulations are obtainable by applying different kernels depending on location \cite{KernelConvolution1996}, \cite{KernelConvolution1998}, \cite{KernelConvolutionFFT}, \cite{CompactCovarianceModelOnASphere}, \cite{KernelConvolutionForRingsOnPlane}, and \cite{FlexibleClassOfKernels}.

\section{Conclusion}

The approach described in this paper extends the applicability of kernel convolution using FFT \cite{KernelConvolutionFFT} to smooth surfaces. The algorithm for finding an integer transformation matrix $A$ is guaranteed to find the optimal one; however, for a more efficient implementation, optimal integer transformation matrices can be precalculated.

This approach is directly applicable to a three-dimensional case where points are located around some smooth surface (sphere, ellipsoid, geoid, etc.).

The algorithms to find an integer point inside an integer triangle and integer tetrahedron are new results. \nameref{sec:FindingAnIntegerPointInsideAnIntegerTetrahedron} describes cases where there is no integer point inside an integer tetrahedron. The open question is, does it cover all cases? This requires further research.

In three dimensions, the optimal integer transformation matrix can have an absolute value of the determinant equal to $ 2 $. This is a new and unexpected result.

\section{Acknowledgment}

The author would like to thank Peter Huggins and Aaron Balog for their helpful discussions and encouragement in the process of writing this paper.

\section*
{
  \label{sec:FindingAnIntegerPointInsideAnIntegerTriangle}
  Appendix I: Finding an Integer Point inside an Integer Triangle
}

In two dimensions, a segment from $\left( a, b \right)$ to $\left( c, d \right)$ has $n - 1$ points inside, where $n = \gcd{\left( c - a, d - b \right)}$.
\begin{equation*}
  \left(
    \dfrac{\left( n - i \right) a + i \cdot c}{n}
    ,
    \dfrac{\left( n - i \right) b + i \cdot d}{n}
  \right)
  ,
  i = \overline{1..n - 1}
  .
\end{equation*}

Pick's theorem \cite{PicksTheoremOriginal} states that the area $A$ of a triangle equals
\begin{equation*}
  A = i + \dfrac{b}{2} - 1
  ,
\end{equation*}
where $i$ is the number of interior points and $b$ is the number of boundary points including three vertices.

Let's consider the case where the triangle does not have any points inside any of its edges. From Pick's theorem and the area of the triangle, the number of points inside the triangle equals $ A - \dfrac{1}{2} $, or from the parallelogram $ \dfrac{P - 1}{2} $, where $ P = 2 A $ is the area of the parallelogram. Therefore, the area of the parallelogram is an odd number. If the triangle has an area that is not equal to $\dfrac{1}{2}$, then it has at least one point inside. For the case where the area equals $\dfrac{3}{2}$, then it has exactly one point inside. Let's divide that triangle by an internal point into three triangles. All of them should not have any points inside or inside their edges. Therefore, they should have areas equal to $\dfrac{1}{2}$. From the fact that the areas of these triangles are proportional to barycentric coordinates, it follows that the coordinates of the internal point are equal to the average of the three vertices of the triangle.

\cite{AnAlgorithmicApproachToPicksTheorem} describes an algorithm for finding a point inside a triangle by analyzing vertices on the line parallel to one of the edges. Another approach, based on triangulation of a matrix, will be discussed. The main advantage is the ability to generalize this approach to higher dimensions.

The next algorithm will be used to find a point inside a triangle, assuming that the triangle has no points inside any of its edges and at least one point inside.
\begin{enumerate}
  \item Find the average of the three vertices of the triangle. If the coordinates are integer numbers, stop.
  \item Round the coordinates. If the point is inside the triangle, stop.
  \item Without loss of generality, assume that one of the triangle vertices is located in the origin of the coordinate system. Let the other two vertices be $\left( a, b \right)$ and $\left( c, d \right)$.
    The point inside the triangle $\left( x, y \right)$ can be represented as a weighted sum of two vectors $\left( a, b \right)$ and $\left( c, d \right)$. Let
    $
      A
      =
      \begin{bmatrix}
        a & c \\
        b & d
      \end{bmatrix}
    $.
    \begin{equation*}
      \begin{bmatrix}
        x \\
        y
      \end{bmatrix}
      =
      A
      \begin{bmatrix}
        \alpha \\
        \beta
      \end{bmatrix}
      ,
    \end{equation*}
    where
    $ 0 < \alpha $, $ 0 < \beta $, and $ \alpha + \beta < 1 $.

    By subtracting, permuting, and changing signs of rows of the matrix $A$, it can always be represented as a product of some matrix $X$, with an absolute value of the determinant equal to $ 1 $, and upper triangular matrix
    \begin{equation}
      R
      =
      \begin{bmatrix}
        1 & m \\
        0 & n
      \end{bmatrix}
      ,
      1 < m < n
      ,
      2 < n
      .
      \label{eq:UpperTriangularMatrixL}
    \end{equation}

    Note that $ m $ and $ n $ are coprime numbers; otherwise, taking weights $ \alpha = 0 $ and $ \beta = \dfrac{1}{\gcd{ \left( m, n \right) }} $ will produce a point on the edge of the triangle. For the same reason, $ m - 1 $ and $ n $ are coprime numbers; otherwise taking  $ \alpha = 1 - \beta $ and $ \beta = \dfrac{1}{\gcd{ \left( m - 1, n \right) }} $ will produce a point on the edge of the triangle. Obviously, $ n $ cannot be an even number.

    Therefore,
    \begin{equation*}
      \begin{bmatrix}
        x \\
        y
      \end{bmatrix}
      =
      X
      R
      \begin{bmatrix}
        \alpha \\
        \beta
      \end{bmatrix}
      .
    \end{equation*}
    If
    $
      R
      \begin{bmatrix}
        \alpha \\
        \beta
      \end{bmatrix}
    $
    is an integer vector, than
    $
      X
      R
      \begin{bmatrix}
        \alpha \\
        \beta
      \end{bmatrix}
    $
    is an integer vector.

    Next, counterexamples will prove that the matrix $ R $ has the form \eqref{eq:UpperTriangularMatrixL}.
    Assume that the matrix $ R $ has the form
    $
      \begin{bmatrix}
        q & m \\
        0 & n
      \end{bmatrix}
    $,
    $ 1 < q $. Then, taking $ \alpha = \dfrac{1}{q} $ and $ \beta = 0 $ will produce a point on the edge of the triangle.
    If the matrix $ R $ has the form
    $
      \begin{bmatrix}
        1 & 0 \\
        0 & 1
      \end{bmatrix}
    $,
    then it does not have any points inside the triangle.
    If the matrix $ R $ has the form
    $
      \begin{bmatrix}
        1 & 0 \\
        0 & n
      \end{bmatrix}
    $,
    then taking $ \alpha = 0 $ and $ \beta = \dfrac{1}{n} $ will produce a point on the edge of the triangle.
    If the matrix $ R $ has the form
    $
      \begin{bmatrix}
        1 & 1 \\
        0 & n
      \end{bmatrix}
    $,
    then taking $ \alpha = \dfrac{n - 1}{n} $ and $ \beta = \dfrac{1}{n} $ will produce a point on the edge of the triangle. This counterexample also proves that $n$ cannot equal $ 2 $. It was also proved before (from Pick's theorem and properties of $ R $), that $n$ cannot be even. This proves that matrix $ R $ satisfies $1 < m < n$ and $2 < n$. Taking $ \alpha = \dfrac{n - m}{n} $ and $ \beta = \dfrac{1}{n} $ will produce a point inside the triangle because $ \alpha + \beta = \dfrac{n - m}{n} + \dfrac{1}{n} < 1 $ and
    $
      R
      \begin{bmatrix}
        \alpha \\
        \beta
      \end{bmatrix}
      =
      \begin{bmatrix}
        1 & m \\
        0 & n
      \end{bmatrix}
      \begin{bmatrix}
        \dfrac{n - m}{n} \\
        \dfrac{1}{n}
      \end{bmatrix}
      =
      \begin{bmatrix}
        1 \\
        1
      \end{bmatrix}
    $,
    which is an integer vector. The point inside is the sum of the columns of the matrix $X$.
\end{enumerate}

Note that iterating over $ \beta = \dfrac{i}{n} $, $ i = \overline{1 .. n - 1} $ will find all points inside the triangle; however, it might not be an efficient approach.

\section*
{
  \label{sec:FindingAnIntegerPointInsideAnIntegerTriangleInHigherDimensions}
  Appendix II: Finding an Integer Point inside an Integer Triangle in Higher Dimensions
}

The approach described in \nameref{sec:FindingAnIntegerPointInsideAnIntegerTriangle} can be applied for triangles in higher dimensions. Assume that there are no points inside the edges of the triangle. The only difference is that the decomposition of rectangular matrix $A$ will have the form
\begin{equation*}
  A
  =
  X
  R
  ,
\end{equation*}
where
\begin{equation*}
  R
  =
  \begin{bmatrix}
    1 & m \\
    0 & n \\
    0 & 0 \\
    \vdots & \vdots \\
    0 & 0
  \end{bmatrix}
  ,
  1 < m < n,
  2 < n
  .
\end{equation*}
$ m $ and $ n $ follow the same set of restrictions as in the two-dimensional case. Taking $ \alpha = \dfrac{n - m}{n} $ and $ \beta = \dfrac{1}{n} $ will produce a point inside the triangle.

\section*
{
  \label{sec:FindingAnIntegerPointInsideAnIntegerTetrahedron}
  Appendix III: Finding an Integer Point inside an Integer Tetrahedron
}

Let's extend the approach described in \nameref{sec:FindingAnIntegerPointInsideAnIntegerTriangle} to a three-dimensional case. Assume that the triangle does not have any points inside its edges or faces (see \nameref{sec:FindingAnIntegerPointInsideAnIntegerTriangleInHigherDimensions}). The matrix $ R $ will have the form
\begin{equation*}
  R
  =
  \begin{bmatrix}
    1 & 0 & a \\
    0 & 1 & b \\
    0 & 0 & n
  \end{bmatrix}
  ,
  0 \leq a < n
  ,
  0 \leq b < n
  .
\end{equation*}
Without loss of generality, assume that $ a \leq b $.

We know $ a \neq 0 $ because the tetrahedron does not have any points on its faces.

Because the tetrahedron does not have any points on its edges, $ a $, $ b $, and $ n $ should satisfy
$
  \gcd
  {
    \left(
      a, b, n
    \right)
  }
  =
  1
$,
$
  \gcd
  {
    \left(
      a - 1, b, n
    \right)
  }
  =
  1
$,
and
$
  \gcd
  {
    \left(
      a, b - 1, n
    \right)
  }
  =
  1
$.
Because this tetrahedron does not have any points on its faces, it should also satisfy
\begin{itemize}
  \item
    $
      \gcd
      {
        \left(
          b, n
        \right)
      }
      =
      1
    $
    for the face formed by
    $ \left( 0, 0, 0 \right) $,
    $ \left( 1, 0, 0 \right) $,
    and
    $ \left( a, b, n \right) $.

  \item
    $
      \gcd
      {
        \left(
          a, n
        \right)
      }
      =
      1
    $
    for the face formed by
    $ \left( 0, 0, 0 \right) $,
    $ \left( 0, 1, 0 \right) $,
    and
    $ \left( a, b, n \right) $.

  \item
    $
      \gcd
      {
        \left(
          a + b - 1, n
        \right)
      }
      =
      1
    $
    for the face formed by
    $ \left( 1, 0, 0 \right) $,
    $ \left( 0, 1, 0 \right) $,
    and
    $ \left( a, b, n \right) $.
\end{itemize}
Note that if conditions for faces not to have any points are satisfied, then there are no points on edges.

The point inside will have the form
\begin{equation*}
  \begin{bmatrix}
    x \\
    y \\
    z
  \end{bmatrix}
  =
  X
  R
  \begin{bmatrix}
    \alpha \\
    \beta \\
    \gamma
  \end{bmatrix}
  ,
\end{equation*}
where
$ 0 < \alpha $, $ 0 < \beta $, $ 0 < \gamma $, and $ \alpha + \beta + \gamma < 1 $.

If $ a = 1 $ or $ n < 4 $, there are no points inside the triangle.

If $ a + b > n + 1 $, then taking $ \alpha = \dfrac{n - a}{n} $, $ \beta = \dfrac{n - b}{n} $ and $ \gamma = \dfrac{1}{n} $ will produce a point inside the triangle.

If $ a + b = n + 1 $, then there are no points inside the triangle, because for all possible weights $ \gamma = \dfrac{i}{n} $, $ i = \overline{1 .. n - 1} $, the sum of all weights is positive and has the form
\begin{equation*}
  \dfrac
  {
    \left( -a \cdot i \right) \!\!\!\!\! \mod n
    +
    \left( -b \cdot i \right) \!\!\!\!\! \mod n
    +
    i
  }
  {
    n
  }
  =
  \dfrac
  {
    \left( -\left( a + b - 1 \right) i \right) \!\!\!\!\! \mod n
  }
  {
    n
  }
  +
  k
  =
  k
  ,
\end{equation*}
where $ k \in \mathbb{N} $. This will contradict the requirement that the sum of all weights is less than $ 1 $.

If $ a + b = n $, from
$
  \gcd
  {
    \left(
      a, n
    \right)
  }
  =
  1
$,
it follows that
\begin{equation*}
  \alpha + \beta
  =
  \dfrac
  {
    \left( -a \cdot i \right) \!\!\!\!\! \mod n
    +
    \left( -b \cdot i \right) \!\!\!\!\! \mod n
  }
  {
    n
  }
  =
  \dfrac
  {
    \left( -a \cdot i \right) \!\!\!\!\! \mod n
    +
    \left( \left( a - n \right) \cdot i \right) \!\!\!\!\! \mod n
  }
  {
    n
  }
  =
  \dfrac
  {
    \left( -a \cdot i \right) \!\!\!\!\! \mod n
    +
    \left( a \cdot i \right) \!\!\!\!\! \mod n
  }
  {
    n
  }
  =
  1
  .
\end{equation*}
Therefore,
$
  \alpha + \beta + \gamma
  =
  1 + \dfrac{i}{n}
  >
  1
$,
which violates the requirement that the sum of all weights is less than $ 1 $.

Experimentally, all other cases, at least for $ n \leq 4096 $, have at least one point inside
\footnote
{
  For $ n \leq 4096 $, if points inside the faces of the tetrahedron are allowed, then the next two tetrahedrons will have no points inside:
  \begin{enumerate}
    \item $ a = 2 $, $ b = 5 $, and $ n = 9 $.
    \item $ a = 3 $, $ b = 5 $, and $ n = 14 $.
  \end{enumerate}
}.
The open question is, are there more tetrahedrons without points inside?

Finding points inside triangles when
$ a + b < n $
is not solved. However, for $ n \leq 4096 $, taking
\begin{equation*}
  \begin{aligned}
    \alpha
    &
    =
    \dfrac{\left( -a \cdot i \right) \!\!\!\!\! \mod n}{n}
    ,
    \\
    \beta
    &
    =
    \dfrac{\left( -b \cdot i \right) \!\!\!\!\! \mod n}{n}
    ,
    \\
    \gamma
    &
    =
    \dfrac{i}{n}
    ,
  \end{aligned}
\end{equation*}
where $ i = \floor {\left( \dfrac{n}{a} \right) } $, when $ \alpha + \beta + \gamma < 1 $ resolves $ 86\% $ of all cases.

Taking
\begin{equation*}
  i
  =
  \left\{
    \begin{aligned}
      & x, & \text{ if } 0 < x,
      \\
      & n + x, & \text{ if } x < 0,
    \end{aligned}
  \right.
\end{equation*}
where $ x $ is found by the Euclidean algorithm for solving $ \left( a + b - 1 \right) x + n \cdot y = 1 $ will solve $ 50\% $ of all cases. $ 19\% $ of all cases are solved by rounded average of all coordinates of the tetrahedron.
$ 7\% $ of all cases are solved by taking $ i = \floor {\left( \dfrac{n}{a} \right) } - 1 $. Combining all these approaches will resolve $ 96.7\% $ of all cases.

\section*
{
  \label{sec:Proof}
  Appendix IV: In Two Dimensions --- The Existence of an Optimal Integer Transformation Matrix\\
  with the Determinant Equal to $ 1 $.
}

Suppose that, for the set of points with a nonempty area of the convex hull, the optimal transformation matrix $A$ has an absolute value of the determinant greater than $ 1 $.

By subtracting, permuting, and changing signs of columns of the matrix $A$, it can always be represented as a product of the lower triangular matrix
\begin{equation*}
  L
  =
  \begin{bmatrix}
    q & 0 \\
    m & n
  \end{bmatrix}
  ,
  0 < q
  ,
  0 \leq m < n
  ,
  1 < n
\end{equation*}
and some matrix $X$, with an absolute value of the determinant equal to $ 1 $. Note that $ q = 1 $ because otherwise, replacing it with $ 1 $ will not make the solution worse. $ m $ and $ n $ are coprime numbers; otherwise, replacing them with $ \dfrac{m}{\gcd{ \left( m, n \right) }} $ and $ \dfrac{n}{\gcd{ \left( m, n \right) }} $ will not make the solution worse. It follows that $ m \neq 0 $.

Therefore,
\begin{equation*}
  L
  =
  \begin{bmatrix}
    1 & 0 \\
    m & n
  \end{bmatrix}
  ,
  0 < m < n
\end{equation*}
and
\begin{equation*}
  \begin{bmatrix}
    x_t \\
    y_t
  \end{bmatrix}
  =
  A
  \begin{bmatrix}
    x \\
    y
  \end{bmatrix}
  =
  L
  X
  \begin{bmatrix}
    x \\
    y
  \end{bmatrix}
  =
  L
  \begin{bmatrix}
    x_r \\
    y_r
  \end{bmatrix}
  ,
\end{equation*}
where $ \left( x, y \right) $ is some point from the original set, $ \left( x_t, y_t \right) $ is a transformed point, and $ \left( x_r, y_r \right) $ is a point after applying the transformation by the matrix $ X $.

Let's consider the worst case scenario for the set of points producing the same minimum volume bounding box as for the set of points $ \left( x_r, y_r \right) $. It will correspond to a parallelogram with sides orthogonal to projection vectors $ \left( 1, 0 \right) $ and $ \left( m, n \right) $. Any point inside the triangle or on its edge between $ \pm \left( 1, 0 \right) $ and $ \pm \left( m, n \right) $ will not have a worse projection than $ \left( 1, 0 \right) $ or $ \left( m, n \right) $. $ \left( 1, 1 \right) $ satisfies such a requirement. Therefore, the matrix $ A $ can be replaced by
\begin{equation*}
  \begin{bmatrix}
    1 & 0 \\
    1 & 1
  \end{bmatrix}
  \text{or}
  \begin{bmatrix}
    1 & 1 \\
    m & n
  \end{bmatrix}
  .
\end{equation*}

Therefore, the determinant of $ A $ is reduced without increasing projections. Recursively applying this operation will produce a determinant equal to $ 1 $. Note that changing signs of the rows of the matrix $ A $ does not change the projections. This proves the existence of an optimal transformation matrix with the determinant equal to $ 1 $.

For any $ 1 < m < n $, the vector $ \left( 1, 1 \right) $ is inside the triangle formed by vectors $ \left( 1, 0 \right) $ and $ \left( m, n \right) $. For any $ m = 1 $ and $ 2 < n $, the vector $ \left( 0, 1 \right) $ is inside the triangle formed by vectors $ \left( -1, 0 \right) $ and $ \left( m, n \right) $. Therefore, it is not possible to have an optimal transformation matrix with a determinant greater than $ 2 $, and, the optimal transformation matrix can only have an absolute value of the determinant equal to $ 1 $ or $ 2 $.

\begin{figure}
  \centering
  \begin{tikzpicture} [scale = 0.9]
    \tkzInit[xmin = -2, ymin = -2, xmax = 2, ymax = 2]
    \tkzGrid

    \draw [->, ultra thick, red] (-1, 0) -- (-1, 1) -- (1, 0) -- (1, -1) -- cycle;

    \tkzAxeXY

    \fill [blue] (0, 0) circle [radius = 3pt];
    \fill [blue] (-1, 0) circle [radius = 3pt];
    \fill [blue] (-1, 1) circle [radius = 3pt];
    \fill [blue] (1, -1) circle [radius = 3pt];
    \fill [blue] (1, 0) circle [radius = 3pt];
  \end{tikzpicture}
  \caption
  {
    Two-dimensional set of points (blue) with convex hull (red), with optimal transformation matrices having absolute values of the determinants equal to $ 1 $ and $ 2 $.
  }
  \label{fig:BadCaseInTwoDimensions}
\end{figure}
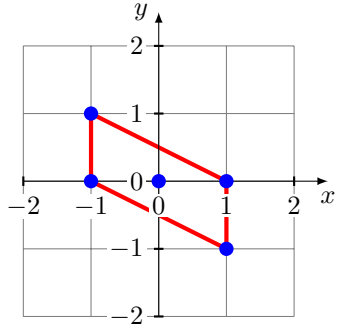

The example of the set of points with the determinant of the optimal transformation matrix
\begin{equation*}
  \begin{bmatrix}
    1 & 0 \\
    1 & 2
  \end{bmatrix}
\end{equation*}
equals $2$:
$ \left( 0, 0 \right) $,
$ \left( -1, 0 \right) $,
$ \left( -1, 1 \right) $,
$ \left( 1, -1 \right) $,
and
$ \left( 1, 0 \right) $,
see Figure~\ref{fig:BadCaseInTwoDimensions}.

However, taking the matrix
\begin{equation*}
  \begin{bmatrix}
    1 & 0 \\
    0 & 1
  \end{bmatrix}
  \text{or}
  \begin{bmatrix}
    1 & 0 \\
    1 & 1
  \end{bmatrix}
\end{equation*}
will produce the same volume for the bounding box.

\section*
{
  \label{sec:Proof3}
  Appendix V: About the Determinant of the Optimal Integer Transformation Matrix in Three Dimensions
}

\begin{figure}
  \centering
  \tdplotsetmaincoords{60}{120}
  \begin{tikzpicture} [tdplot_main_coords, line/.style={very thick, black}, convex hull line/.style={very thick, red}, point/.style={ball color = blue}, convex hull point/.style={ball color = red}]
    \draw [convex hull line] (-2, -2, 1) -- (-2, -2, 3);
    \draw [convex hull line] (-2, -2, 1) -- (-2, 2, -1);
    \draw [convex hull line] (-2, -2, 1) -- (2, -2, -1);
    \draw [convex hull line] (-2, -2, 3) -- (-2, 2, 1);
    \draw [convex hull line] (-2, -2, 3) -- (2, -2, 1);

    \draw [line] (2, 2, -3) -- (2, 2, -2);
    \draw [line] (2, 2, -2) -- (2, 2, -1);
    \draw [line] (2, 1, -2) -- (2, 2, -2);
    \draw [line] (2, 1, -2) -- (2, 1, -1);
    \draw [line] (2, 1, -1) -- (2, 2, -1);
    \draw [line] (2, 0, -2) -- (2, 1, -2);
    \draw [line] (2, 0, -2) -- (2, 0, -1);
    \draw [line] (2, 0, -1) -- (2, 1, -1);
    \draw [line] (2, 0, -1) -- (2, 0, 0);
    \draw [line] (2, -1, -1) -- (2, 0, -1);
    \draw [line] (2, -1, -1) -- (2, -1, 0);
    \draw [line] (2, -1, 0) -- (2, 0, 0);
    \draw [line] (2, -2, -1) -- (2, -1, -1);
    \draw [line] (2, -2, -1) -- (2, -2, 0);
    \draw [line] (2, -2, 0) -- (2, -1, 0);
    \draw [line] (2, -2, 0) -- (2, -2, 1);
    \draw [line] (1, 2, -2) -- (2, 2, -2);
    \draw [line] (1, 2, -2) -- (1, 2, -1);
    \draw [line] (1, 2, -1) -- (2, 2, -1);
    \draw [line] (1, 1, -2) -- (2, 1, -2);
    \draw [line] (1, 1, -2) -- (1, 2, -2);
    \draw [line] (1, 1, -2) -- (1, 1, -1);
    \draw [line] (1, 1, -1) -- (2, 1, -1);
    \draw [line] (1, 1, -1) -- (1, 2, -1);
    \draw [line] (1, 1, -1) -- (1, 1, 0);
    \draw [line] (1, 0, -1) -- (2, 0, -1);
    \draw [line] (1, 0, -1) -- (1, 1, -1);
    \draw [line] (1, 0, -1) -- (1, 0, 0);
    \draw [line] (1, 0, 0) -- (2, 0, 0);
    \draw [line] (1, 0, 0) -- (1, 1, 0);
    \draw [line] (1, -1, -1) -- (2, -1, -1);
    \draw [line] (1, -1, -1) -- (1, 0, -1);
    \draw [line] (1, -1, -1) -- (1, -1, 0);
    \draw [line] (1, -1, 0) -- (2, -1, 0);
    \draw [line] (1, -1, 0) -- (1, 0, 0);
    \draw [line] (1, -1, 0) -- (1, -1, 1);
    \draw [line] (1, -2, 0) -- (2, -2, 0);
    \draw [line] (1, -2, 0) -- (1, -1, 0);
    \draw [line] (1, -2, 0) -- (1, -2, 1);
    \draw [line] (1, -2, 1) -- (2, -2, 1);
    \draw [line] (1, -2, 1) -- (1, -1, 1);
    \draw [line] (0, 2, -2) -- (1, 2, -2);
    \draw [line] (0, 2, -2) -- (0, 2, -1);
    \draw [line] (0, 2, -1) -- (1, 2, -1);
    \draw [line] (0, 2, -1) -- (0, 2, 0);
    \draw [line] (0, 1, -1) -- (1, 1, -1);
    \draw [line] (0, 1, -1) -- (0, 2, -1);
    \draw [line] (0, 1, -1) -- (0, 1, 0);
    \draw [line] (0, 1, 0) -- (1, 1, 0);
    \draw [line] (0, 1, 0) -- (0, 2, 0);
    \draw [line] (0, 0, -1) -- (1, 0, -1);
    \draw [line] (0, 0, -1) -- (0, 1, -1);
    \draw [line] (0, 0, -1) -- (0, 0, 0);
    \draw [line] (0, 0, 0) -- (1, 0, 0);
    \draw [line] (0, 0, 0) -- (0, 1, 0);
    \draw [line] (0, 0, 0) -- (0, 0, 1);
    \draw [line] (0, -1, 0) -- (1, -1, 0);
    \draw [line] (0, -1, 0) -- (0, 0, 0);
    \draw [line] (0, -1, 0) -- (0, -1, 1);
    \draw [line] (0, -1, 1) -- (1, -1, 1);
    \draw [line] (0, -1, 1) -- (0, 0, 1);
    \draw [line] (0, -2, 0) -- (1, -2, 0);
    \draw [line] (0, -2, 0) -- (0, -1, 0);
    \draw [line] (0, -2, 0) -- (0, -2, 1);
    \draw [line] (0, -2, 1) -- (1, -2, 1);
    \draw [line] (0, -2, 1) -- (0, -1, 1);
    \draw [line] (0, -2, 1) -- (0, -2, 2);
    \draw [line] (-1, 2, -1) -- (0, 2, -1);
    \draw [line] (-1, 2, -1) -- (-1, 2, 0);
    \draw [line] (-1, 2, 0) -- (0, 2, 0);
    \draw [line] (-1, 1, -1) -- (0, 1, -1);
    \draw [line] (-1, 1, -1) -- (-1, 2, -1);
    \draw [line] (-1, 1, -1) -- (-1, 1, 0);
    \draw [line] (-1, 1, 0) -- (0, 1, 0);
    \draw [line] (-1, 1, 0) -- (-1, 2, 0);
    \draw [line] (-1, 1, 0) -- (-1, 1, 1);
    \draw [line] (-1, 0, 0) -- (0, 0, 0);
    \draw [line] (-1, 0, 0) -- (-1, 1, 0);
    \draw [line] (-1, 0, 0) -- (-1, 0, 1);
    \draw [line] (-1, 0, 1) -- (0, 0, 1);
    \draw [line] (-1, 0, 1) -- (-1, 1, 1);
    \draw [line] (-1, -1, 0) -- (0, -1, 0);
    \draw [line] (-1, -1, 0) -- (-1, 0, 0);
    \draw [line] (-1, -1, 0) -- (-1, -1, 1);
    \draw [line] (-1, -1, 1) -- (0, -1, 1);
    \draw [line] (-1, -1, 1) -- (-1, 0, 1);
    \draw [line] (-1, -1, 1) -- (-1, -1, 2);
    \draw [line] (-1, -2, 1) -- (0, -2, 1);
    \draw [line] (-1, -2, 1) -- (-1, -1, 1);
    \draw [line] (-1, -2, 1) -- (-1, -2, 2);
    \draw [line] (-1, -2, 2) -- (0, -2, 2);
    \draw [line] (-1, -2, 2) -- (-1, -1, 2);
    \draw [line] (-2, 2, -1) -- (-1, 2, -1);
    \draw [line] (-2, 2, -1) -- (-2, 2, 0);
    \draw [line] (-2, 2, 0) -- (-1, 2, 0);
    \draw [line] (-2, 2, 0) -- (-2, 2, 1);
    \draw [line] (-2, 1, 0) -- (-1, 1, 0);
    \draw [line] (-2, 1, 0) -- (-2, 2, 0);
    \draw [line] (-2, 1, 0) -- (-2, 1, 1);
    \draw [line] (-2, 1, 1) -- (-1, 1, 1);
    \draw [line] (-2, 1, 1) -- (-2, 2, 1);
    \draw [line] (-2, 0, 0) -- (-1, 0, 0);
    \draw [line] (-2, 0, 0) -- (-2, 1, 0);
    \draw [line] (-2, 0, 0) -- (-2, 0, 1);
    \draw [line] (-2, 0, 1) -- (-1, 0, 1);
    \draw [line] (-2, 0, 1) -- (-2, 1, 1);
    \draw [line] (-2, 0, 1) -- (-2, 0, 2);
    \draw [line] (-2, -1, 1) -- (-1, -1, 1);
    \draw [line] (-2, -1, 1) -- (-2, 0, 1);
    \draw [line] (-2, -1, 1) -- (-2, -1, 2);
    \draw [line] (-2, -1, 2) -- (-1, -1, 2);
    \draw [line] (-2, -1, 2) -- (-2, 0, 2);
    \draw [line] (-2, -2, 1) -- (-1, -2, 1);
    \draw [line] (-2, -2, 1) -- (-2, -1, 1);
    \draw [line] (-2, -2, 1) -- (-2, -2, 2);
    \draw [line] (-2, -2, 2) -- (-1, -2, 2);
    \draw [line] (-2, -2, 2) -- (-2, -1, 2);
    \draw [line] (-2, -2, 2) -- (-2, -2, 3);

    \draw [convex hull line] (2, 2, -1) -- (2, 2, -3);
    \draw [convex hull line] (2, 2, -1) -- (2, -2, 1);
    \draw [convex hull line] (2, 2, -1) -- (-2, 2, 1);
    \draw [convex hull line] (2, 2, -3) -- (2, -2, -1);
    \draw [convex hull line] (2, 2, -3) -- (-2, 2, -1);
    \draw [convex hull line] (-2, 2, -1) -- (-2, 2, 1);
    \draw [convex hull line] (2, -2, -1) -- (2, -2, 1);

    \shade [point] (2, 2, -3) circle (0.1 cm);
    \shade [point] (2, 2, -2) circle (0.1 cm);
    \shade [point] (2, 2, -1) circle (0.1 cm);
    \shade [point] (2, 1, -2) circle (0.1 cm);
    \shade [point] (2, 1, -1) circle (0.1 cm);
    \shade [point] (2, 0, -2) circle (0.1 cm);
    \shade [point] (2, 0, -1) circle (0.1 cm);
    \shade [point] (2, 0, 0) circle (0.1 cm);
    \shade [point] (2, -1, -1) circle (0.1 cm);
    \shade [point] (2, -1, 0) circle (0.1 cm);
    \shade [point] (2, -2, -1) circle (0.1 cm);
    \shade [point] (2, -2, 0) circle (0.1 cm);
    \shade [point] (2, -2, 1) circle (0.1 cm);
    \shade [point] (1, 2, -2) circle (0.1 cm);
    \shade [point] (1, 2, -1) circle (0.1 cm);
    \shade [point] (1, 1, -2) circle (0.1 cm);
    \shade [point] (1, 1, -1) circle (0.1 cm);
    \shade [point] (1, 1, 0) circle (0.1 cm);
    \shade [point] (1, 0, -1) circle (0.1 cm);
    \shade [point] (1, 0, 0) circle (0.1 cm);
    \shade [point] (1, -1, -1) circle (0.1 cm);
    \shade [point] (1, -1, 0) circle (0.1 cm);
    \shade [point] (1, -1, 1) circle (0.1 cm);
    \shade [point] (1, -2, 0) circle (0.1 cm);
    \shade [point] (1, -2, 1) circle (0.1 cm);
    \shade [point] (0, 2, -2) circle (0.1 cm);
    \shade [point] (0, 2, -1) circle (0.1 cm);
    \shade [point] (0, 2, 0) circle (0.1 cm);
    \shade [point] (0, 1, -1) circle (0.1 cm);
    \shade [point] (0, 1, 0) circle (0.1 cm);
    \shade [point] (0, 0, -1) circle (0.1 cm);
    \shade [point] (0, 0, 0) circle (0.1 cm);
    \shade [point] (0, 0, 1) circle (0.1 cm);
    \shade [point] (0, -1, 0) circle (0.1 cm);
    \shade [point] (0, -1, 1) circle (0.1 cm);
    \shade [point] (0, -2, 0) circle (0.1 cm);
    \shade [point] (0, -2, 1) circle (0.1 cm);
    \shade [point] (0, -2, 2) circle (0.1 cm);
    \shade [point] (-1, 2, -1) circle (0.1 cm);
    \shade [point] (-1, 2, 0) circle (0.1 cm);
    \shade [point] (-1, 1, -1) circle (0.1 cm);
    \shade [point] (-1, 1, 0) circle (0.1 cm);
    \shade [point] (-1, 1, 1) circle (0.1 cm);
    \shade [point] (-1, 0, 0) circle (0.1 cm);
    \shade [point] (-1, 0, 1) circle (0.1 cm);
    \shade [point] (-1, -1, 0) circle (0.1 cm);
    \shade [point] (-1, -1, 1) circle (0.1 cm);
    \shade [point] (-1, -1, 2) circle (0.1 cm);
    \shade [point] (-1, -2, 1) circle (0.1 cm);
    \shade [point] (-1, -2, 2) circle (0.1 cm);
    \shade [point] (-2, 2, -1) circle (0.1 cm);
    \shade [point] (-2, 2, 0) circle (0.1 cm);
    \shade [point] (-2, 2, 1) circle (0.1 cm);
    \shade [point] (-2, 1, 0) circle (0.1 cm);
    \shade [point] (-2, 1, 1) circle (0.1 cm);
    \shade [point] (-2, 0, 0) circle (0.1 cm);
    \shade [point] (-2, 0, 1) circle (0.1 cm);
    \shade [point] (-2, 0, 2) circle (0.1 cm);
    \shade [point] (-2, -1, 1) circle (0.1 cm);
    \shade [point] (-2, -1, 2) circle (0.1 cm);
    \shade [point] (-2, -2, 1) circle (0.1 cm);
    \shade [point] (-2, -2, 2) circle (0.1 cm);
    \shade [point] (-2, -2, 3) circle (0.1 cm);

    \shade [convex hull point] (-2, -2, 1) circle (0.1 cm);
    \shade [convex hull point] (-2, -2, 3) circle (0.1 cm);
    \shade [convex hull point] (-2, 2, -1) circle (0.1 cm);
    \shade [convex hull point] (-2, 2, 1) circle (0.1 cm);
    \shade [convex hull point] (2, -2, -1) circle (0.1 cm);
    \shade [convex hull point] (2, -2, 1) circle (0.1 cm);
    \shade [convex hull point] (2, 2, -3) circle (0.1 cm);
    \shade [convex hull point] (2, 2, -1) circle (0.1 cm);
  \end{tikzpicture}
  \caption
  {
    Three-dimensional set of points (blue) with convex hull vertices and edges (red) in an integer grid (black) with the optimal transformation matrix having an absolute value of the determinant equal to $ 2 $.
  }
  \label{fig:BadCaseInThreeDimensions}
\end{figure}
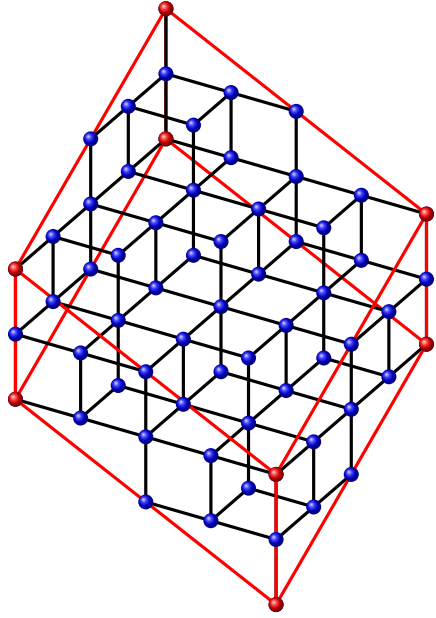

From the previous appendix, the proof that there is an optimal transformation matrix with the determinant equal to $ 1 $ is not valid in three dimensions. Here is an example of the set of points (see Figure~\ref{fig:BadCaseInThreeDimensions}) without an optimal transformation matrix with an absolute value of the determinant equal to $ 1 $. The vertices of the convex hull:
$ \left( -2, -2, 1\right) $,
$ \left( -2, -2, 3\right) $,
$ \left( -2, 2, -1\right) $,
$ \left( -2, 2, 1\right) $,
$ \left( 2, -2, -1\right) $,
$ \left( 2, -2, 1\right) $,
$ \left( 2, 2, -3\right) $,
and
$ \left( 2, 2, -1\right) $.
The complete list of points:
$ \left( -2, -2, 1 \right) $,
$ \left( -2, -2, 2 \right) $,
$ \left( -2, -2, 3 \right) $,
$ \left( -2, -1, 1 \right) $,
$ \left( -2, -1, 2 \right) $,
$ \left( -2, 0, 0 \right) $,
$ \left( -2, 0, 1 \right) $,
$ \left( -2, 0, 2 \right) $,
$ \left( -2, 1, 0 \right) $,
$ \left( -2, 1, 1 \right) $,
$ \left( -2, 2, -1 \right) $,
$ \left( -2, 2, 0 \right) $,
$ \left( -2, 2, 1 \right) $,
$ \left( -1, -2, 1 \right) $,
$ \left( -1, -2, 2 \right) $,
$ \left( -1, -1, 0 \right) $,
$ \left( -1, -1, 1 \right) $,
$ \left( -1, -1, 2 \right) $,
$ \left( -1, 0, 0 \right) $,
$ \left( -1, 0, 1 \right) $,
$ \left( -1, 1, -1 \right) $,
$ \left( -1, 1, 0 \right) $,
$ \left( -1, 1, 1 \right) $,
$ \left( -1, 2, -1 \right) $,
$ \left( -1, 2, 0 \right) $,
$ \left( 0, -2, 0 \right) $,
$ \left( 0, -2, 1 \right) $,
$ \left( 0, -2, 2 \right) $,
$ \left( 0, -1, 0 \right) $,
$ \left( 0, -1, 1 \right) $,
$ \left( 0, 0, -1 \right) $,
$ \left( 0, 0, 0 \right) $,
$ \left( 0, 0, 1 \right) $,
$ \left( 0, 1, -1 \right) $,
$ \left( 0, 1, 0 \right) $,
$ \left( 0, 2, -2 \right) $,
$ \left( 0, 2, -1 \right) $,
$ \left( 0, 2, 0 \right) $,
$ \left( 1, -2, 0 \right) $,
$ \left( 1, -2, 1 \right) $,
$ \left( 1, -1, -1 \right) $,
$ \left( 1, -1, 0 \right) $,
$ \left( 1, -1, 1 \right) $,
$ \left( 1, 0, -1 \right) $,
$ \left( 1, 0, 0 \right) $,
$ \left( 1, 1, -2 \right) $,
$ \left( 1, 1, -1 \right) $,
$ \left( 1, 1, 0 \right) $,
$ \left( 1, 2, -2 \right) $,
$ \left( 1, 2, -1 \right) $,
$ \left( 2, -2, -1 \right) $,
$ \left( 2, -2, 0 \right) $,
$ \left( 2, -2, 1 \right) $,
$ \left( 2, -1, -1 \right) $,
$ \left( 2, -1, 0 \right) $,
$ \left( 2, 0, -2 \right) $,
$ \left( 2, 0, -1 \right) $,
$ \left( 2, 0, 0 \right) $,
$ \left( 2, 1, -2 \right) $,
$ \left( 2, 1, -1 \right) $,
$ \left( 2, 2, -3 \right) $,
$ \left( 2, 2, -2 \right) $,
and
$ \left( 2, 2, -1 \right) $.
The optimal transformation matrix
\begin{equation*}
  \begin{bmatrix}
    1 & 0 & 0 \\
    0 & 1 & 0 \\
    1 & 1 & 2
  \end{bmatrix}
  .
\end{equation*}

Proof of the existence of an optimal transformation matrix with an absolute value of the determinant equal to $ 1 $ or $ 2 $ is similar to one described in \nameref{sec:Proof}. Following this proof, the lower triangular matrix $ L $ will have the form
\begin{equation*}
  \begin{bmatrix}
    q & 0 & 0 \\
    m & n & 0 \\
    * & * & *
  \end{bmatrix}
  ,
  0 < q
  ,
  0 \leq m < n
  ,
\end{equation*}
where $ * $ is some value.

From \nameref{sec:Proof}, it follows that $ q = 1 $, and it is always possible to replace projection vectors to have $ n = 1 $. Therefore, matrix $ L $ can always be represented in the form
\begin{equation*}
  \begin{bmatrix}
    1 & 0 & 0 \\
    0 & 1 & 0 \\
    a & b & n
  \end{bmatrix}
  ,
  0 \leq a \leq b < n
  .
\end{equation*}

Similar to the approach described in \nameref{sec:FindingAnIntegerPointInsideAnIntegerTriangleInHigherDimensions}, a vector inside, on the edge, or on the face one of the tetrahedrons formed by the next three vectors $ \pm \left( 1, 0, 0 \right) $, $ \pm \left( 0, 1, 0 \right) $, and $\left( a, b, n \right) $ is found by finding a set of $ \alpha $, $ \beta $, and $ \gamma $ satisfying $ 0 < \left| \alpha \right| + \left| \beta \right| + \left| \gamma \right| \leq 1 $, $ \left| \alpha \right| \neq 1 $, $ \left| \beta \right| \neq 1 $, $ \left| \gamma \right| \neq 1 $ and producing the vector
\begin{equation*}
  \begin{bmatrix}
    1 & 0 & a \\
    0 & 1 & b \\
    0 & 0 & n
  \end{bmatrix}
  \begin{bmatrix}
    \alpha \\
    \beta \\
    \gamma
  \end{bmatrix}
  .
\end{equation*}
Taking
\begin{equation*}
  \alpha
  =
  \left\{
    \begin{aligned}
      -\dfrac{a}{n}, a < n - a,  \\
      \dfrac{n - a}{n}, \text{otherwise},
    \end{aligned}
  \right.
\end{equation*}
\begin{equation*}
  \beta
  =
  \left\{
    \begin{aligned}
      -\dfrac{b}{n}, b < n - b,  \\
      \dfrac{n - b}{n}, \text{otherwise},
    \end{aligned}
  \right.
\end{equation*}
\begin{equation*}
  \gamma
  =
  \dfrac{1}{n}
\end{equation*}
will have
$ \left| \alpha \right| + \left| \beta \right| + \left| \gamma \right| > 1 $ only in the case where $ 2 a = n $ and $ 2 b = n $. Unless $ n = 2 $, it can be replace by $ a = 1 $, $ b = 1 $, and $ n = 2 $, to produce a better solution. This proves the existence of an optimal transformation matrix with a determinant equal to $ 1 $ or $ 2 $.

The open question is, what is the possible absolute value of the determinant of an optimal transformation matrix?


\newcommand{\doi}[1]{\textsc{doi}: \href{http://dx.doi.org/#1}{\nolinkurl{#1}}}


\begingroup
\raggedright
\bibliographystyle{IEEEtran}
\bibliography{EfficientKernelConvolutionForSmoothSurfaces}
\endgroup


\end{document}